# Observation of hydrostatic-pressure-modulated giant caloric effect and electronic topological transition

Jinying Yang†, Xingchen Liu†, Yibo Wang, Shen Zhang, Yang Liu, Xuebin Dong, Yiting Feng, Qiusa Ren, Ping He, Meng Lyu, Binbin Wang, Shouguo Wang, Guangheng Wu, Xixiang Zhang* and Enke Liu*

Jinying Yang, Xingchen Liu, Yibo Wang, Shen Zhang, Yang Liu, Xuebin Dong, Yiting Feng, Ping He, Meng Lyu, Binbin Wang, Guangheng Wu, Enke Liu
Beijing National Laboratory for Condensed Matter Physics, Institute of Physics, Chinese Academy of Sciences, Beijing 100190, China
E-mail: ekliu@iphy.ac.cn

Jinying Yang, Xingchen Liu, Yibo Wang, Shen Zhang, Yang Liu, Xuebin Dong, Yiting Feng, Ping He
University of Chinese Academy of Sciences, Beijing 100049, China

Qiusa Ren, Shouguo Wang
School of Materials Science and Engineering, University of Science and Technology Beijing, Beijing, 100083, China

Shouguo Wang
Anhui Key Laboratory of Magnetic Functional Materials and Devices, School of Materials Science and Engineering, Anhui University, Hefei 230601, China

Xixiang Zhang
Physical Science and Engineering Division (PSE), King Abdullah University of Science and Technology (KAUST), Thuwal 23955-6900, Saudi Arabia
E-mail: xixiang.zhang@kaust.edu.sa

## Abstract

Phase transition is a fundamental phenomenon in condensed matter physics, in which states of matter transform to each other with various critical behaviors under different conditions. The magnetic martensitic transformation features significant multi-caloric effects that benefit the solid-state cooling or heat pumping. Meanwhile, the electronic topological transition (ETT) driven by pressure has been rarely reported in martensitic systems. Here, the modulation effects of hydrostatic pressure on phase transitions in a magnetic martensitic alloy are reported. Owing to the huge volume expansion during the transition, the martensitic transition temperature is driven from 339 to 273 K by pressure within 1 GPa, resulting in highly tunable giant baro- and magneto-caloric effects (BCE and MCE) in a wide working temperature range. Interestingly, an ETT was further induced by pressure in the martensite phase, with a sudden drop of the measured saturation magnetization around 0.6 GPa. First-principles calculations reveal a sharp change in the density of states (DOS) due to the orbit shift around the Fermi level at the same pressure and reproduce the experimental observation of magnetization. Besides, the ETT is accompanied by remarkable changes in the lattice parameters and the unit-cell orthorhombicity. The study provides insight into pressure-modulated exotic phase-transition phenomena in magnetic martensitic systems.

# 1. Introduction

MMX (M represents the transition group elements, and X the main group elements) alloys are a family with martensitic phase transition (MPT).[1] The features of environmental friendliness, low cost of raw materials, and wide phase transition temperature range make MMX a good candidate for solid-state refrigeration.[2] However, the MPT temperatures of undoped MMX alloys are higher than room temperature, and even above 1000 K for Si - based MMX alloys.[3] For Ge - based compounds such as MnNiGe (483 K) and MnCoGe (420 K), the MPT temperatures are slightly higher than room temperature.[4] Therefore, most studies have focused on doped MnNiGe and MnCoGe under pressure to obtain the continuously tunable magnetic MPT near room temperature.[5,6] Changes in magnetic properties (paramagnetic to ferromagnetic), crystal structure (hexagonal to tetragonal), and volume (volume expansion) occur around the MPTs. Therefore, magnetic field, chemical doping, and pressure are the efficient ways to tune the phase transition of MMX alloys.[2,7-13] Previous studies indicated that high modulation efficiency (≈-100 K/GPa) demonstrates the effectiveness and tunability of pressure regulation on the giant caloric effects of MMX alloys.[5-10,14] Some other changes also appear under pressure, such as the transition from spiral antiferromagnetism to ferromagnetism, exchange bias, etc.[14,15]

It is known that the pressure as well as the electric field, temperature and impurity is an effective degree of freedom to change the topology of the Fermi surface. This phenomenon is known as the ETT or the Lifshitz transition, which corresponds to the disruption of the "neck" of the Fermi surface, the appearance or disappearance of electronic pockets.[16-22] The band structure and DOS around the Fermi level are directly related to the changes in the Fermi surface. During the ETT, the saturation magnetization, carrier type and carrier concentration of the system have a potential change due to the changes in the DOS at the Fermi level. An ETT induced by hydrostatic pressure has not been reported in MMX alloys, which is involved in our present work.

$Mn_{0.54}Fe_{0.46}NiGe_{0.4}Si_{0.6}$, bearing a MPT close to the room temperature, is one of the compositions of $Mn_{1-y}Fe_yNiGe_{1-x}Si_x$ (0.26 ≤ y ≤ 0.55, 0 ≤ x ≤ 1) family developed by our group, which is a MPT system with extremely wide Curie-temperature windows (405 K) and large volume expansions (2.5 - 3.6%).[23] These advantages endue this material with the potential for high pressure driving efficiency and a large modulation range in temperature. In this work, hydrostatic pressure was used to modulate the MPT and magnetic properties. The large adjustable caloric effect in a wide temperature range was obtained under hydrostatic pressure and magnetic field. It provides a material platform for solid-state refrigeration controlled in multiple ways. A sudden decrease in saturation magnetization ($M_S$) occurred ≈0.6 GPa, indicating an ETT here, and it is accompanied by remarkable decreases in both c axis and spin polarization of the system.

# 2. Results and Discussion

## 2.1. Pressure Driven Giant Baro-Caloric Effect

Mn$_{0.54}$Fe$_{0.46}$NiGe$_{0.4}$Si$_{0.6}$ crystallizes in a Ni$_2$In-type hexagonal structure (*P*6$_3$/*mmc*, space group no. 194) (**Figure 1**a). Ni and Ge/Si form the layered structure of six-membered rings, and Mn/Fe is interlayered. At 339 K, the MPT occurs and the crystal structure transforms into a TiNiSi-type orthorhombic structure (*Pnma*, space group no. 62).[1,3,4,24] Figure 1b shows that the six-membered rings are distorted along the direction of *c* axis, forming an arm-chair structure. Mn/Fe atoms are thus arranged in zigzag lines instead of straight lines. Figure 1c demonstrates that in the martensitic phase, the magnetic moment is primarily carried by Mn/Fe.[23] There is a strong electron localization between Ge/Si and Ni, which indicates the covalent bond between Ge/Si-Ni. The metallic bond between Mn-Mn is also observed in Figure 1d.[1] The volume expansion during the MPT is 2.8%,[5,23] which is very large compared to other martensitic transition materials[25,26] and is beneficial for modulating pressure-driven MPT.

The MPT of Mn$_{0.54}$Fe$_{0.46}$NiGe$_{0.4}$Si$_{0.6}$ is accompanied by the changes in structure and magnetic properties. We measured the magnetostructural transition under hydrostatic pressure in a magnetic field of 0.01 T. Figure 1e shows the MPT from the paramagnetic parent phase to the ferromagnetic martensitic phase at the temperature of 339 K under zero pressure. The observed temperature hysteresis indicates that this is the first order martensitic structural transition. Since the volume expands (2.8%) during the phase transition, pressure is also a way to modulate the MPT. The magnetostructural transition was tuned from 339 to 273 K within 1 GPa, which shows that the pressure can effectively modulate the phase transition with a huge volume expansion. For the volume-expansion MPT, the stabilizing effect of hydrostatic pressure on the parent phase is generally observed in the MMX alloys.[5,6,8,14] But the MPT temperature of some materials is close to the Curie temperature of the parent phase, which was decreased out of the Curie-temperature window by pressure, leading to a rapid decrease in magnetic entropy change.[5-11,14,27]

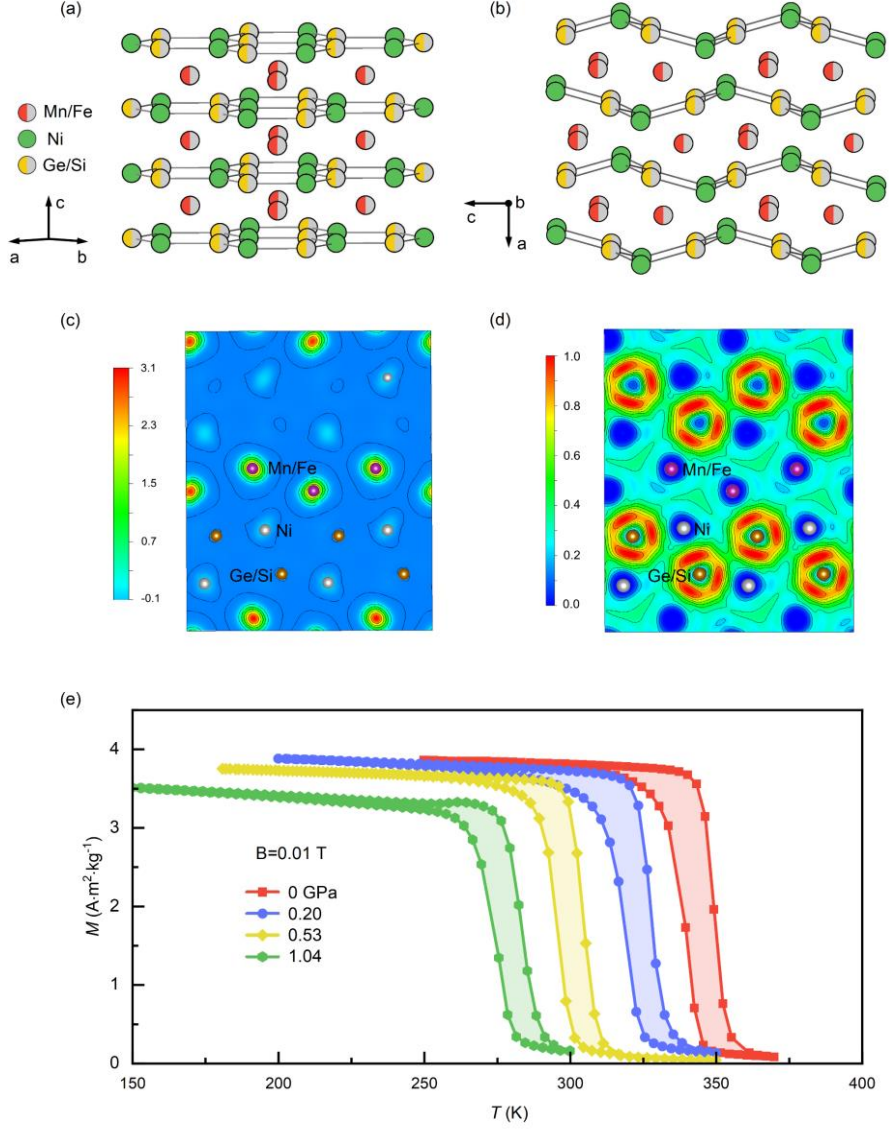

**Figure 1.** Panels (a) and (b) show the crystal structure of $Mn_{0.54}Fe_{0.46}NiGe_{0.4}Si_{0.6}$ parent phase and martensitic phase, respectively. c,d) Contours of spin electron density (SPD) and valence-electron localization function (ELF) of martensitic phase of $Mn_{0.5}Fe_{0.5}NiGe_{0.5}Si_{0.5}$. The crystal orientation is consistent with (b). e) Magnetostructural transition measured at a magnetic field of 0.01 T and tuned by hydrostatic pressure.

The $T_t$-$P$ curve obtained from Figure 1e was fitted with linear and parabolic forms, respectively (**Figure 2**). The Clausius–Clapeyron equation $\Delta S_t = V\Delta\omega(\frac{dT_t}{dP})^{-1}$ was used to calculate the structural entropy change.[6] Here, $V\Delta\omega$ is the volume difference between the martensitic and parent phases. For these fitting, the obtained -$\Delta S_t$ values are 61.9 J·K$^{-1}$·kg$^{-1}$ (linear fitting) and 36.5 J·K$^{-1}$·kg$^{-1}$ (parabolic fitting). It indicates that the linear fitting is more reasonable as the -$\Delta S_t$ value obtained by the linear fitting is closer to the entropy change value obtained by the differential scanning calorimeter in our previous report (55.6 J·K$^{-1}$·kg$^{-1}$)[23]. At low hydrostatic pressures, a constant -$\Delta S_t$ can be expected since $V\Delta\omega$ and $dT_t/dP$ remain unchanged. A giant BCE can be expected in these materials, owing to the giant volume change of 2.8%, which is comparable to the $-\Delta S_t$ of the $(MnNiSi)_{0.46}(MnFeGe)_{0.54}$[6] and $(MnNiSi)_{0.6}(FeCoGe)_{0.4}$[11].

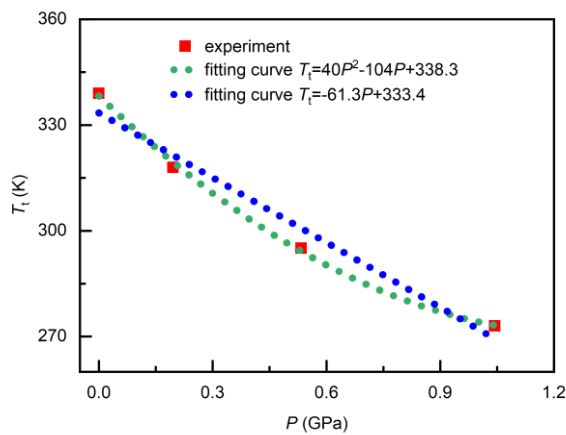

**Figure 2.** The MPT temperature driven by the hydrostatic pressure. The data of red squares are obtained from the experimental data in Figure 1e. And the green and blue dashed lines are the fitting curves in parabolic and linear forms, respectively.

The loop method was employed to measure the magnetic entropy change.[28] The magnetization curves of the $Mn_{0.54}Fe_{0.46}NiGe_{0.4}Si_{0.6}$ that were measured during cooling (**Figure 3**a,b). The curves of Figure 3a,b were measured between 356 to 332 K under zero pressure and 294 to 268 K under 1.04 GPa, respectively. At high temperatures, the magnetization curves of the parent phase are straight lines, showing the process of the paramagnetic magnetization. With the temperature decreases, the metamagnetic transition behavior can be observed during the magnetization process, which corresponds to the magnetic field induced martensitic transition. As the temperature decreases further, the sample enters the ferromagnetic martensite phase completely. A typical ferromagnetic magnetization can be thus observed. Figure 3c displays the magnetic entropy change ($-\Delta S_m$) under pressure that was calculated by Maxwell's relation. The temperature of maximum magnetic entropy corresponds to the temperature of the martensitic phase transition in Figures 1e and 2. The $-\Delta S_m$ increases with the increase of the magnetic field. Figure 3c shows that the magnetic phase transition was continuously driven to lower temperatures by hydrostatic pressure. And the large MCE of $\approx -30$ J·K$^{-1}$·kg$^{-1}$ (5 T) can be tuned from 339 to 273 K within 1 GPa. The $-\Delta S_m$ of magnetic MPT (paramagnetic parent phase to ferromagnetic martensite phase) of MMX is almost a constant under moderate pressures, indicating that it is robust under pressure.[8-11] The volume-expansion martensitic transformation with a large MCE can be tuned by pressure, which makes $Mn_{0.54}Fe_{0.46}NiGe_{0.4}Si_{0.6}$ a promising magnetic refrigerant material with a wide working temperature range.[9,14]

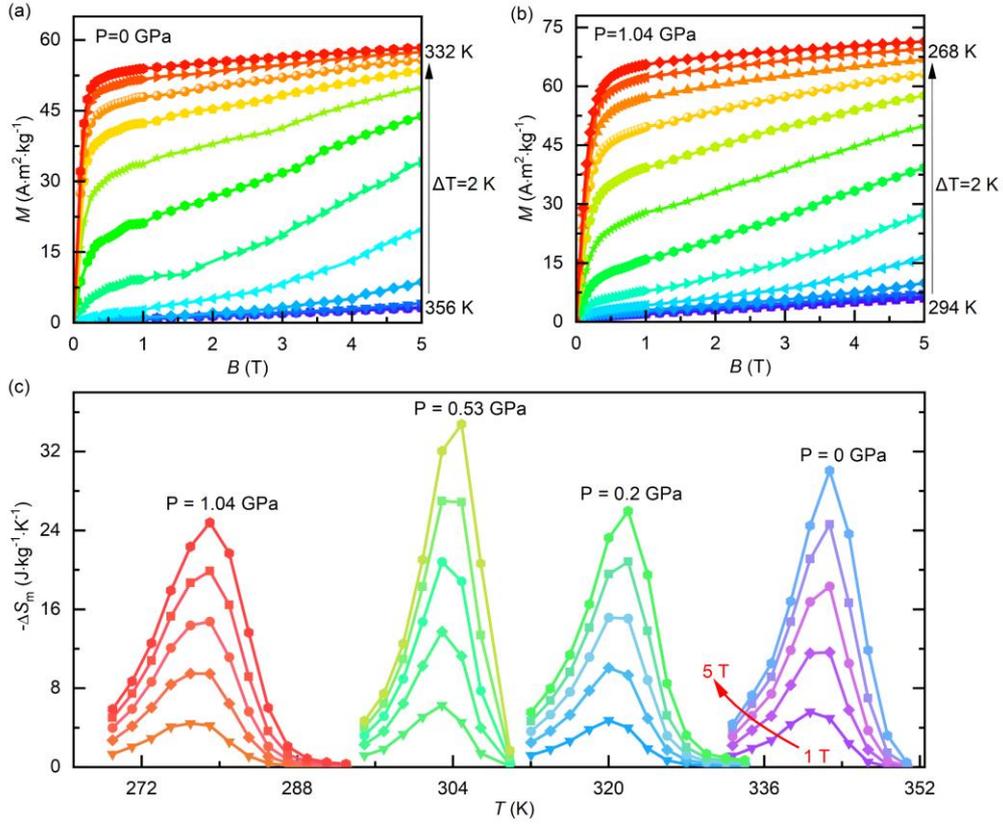

**Figure 3.** MCE tuned by hydrostatic pressure. Panels (a) and (b) show the magnetization curves at different temperatures. Panel (a) shows the magnetization curves under 0 GPa. Panel (b) shows the magnetization curves under 1.04 GPa. c) The magnetic entropy changes in a magnetic field change from 1 to 5 T under hydrostatic pressure.

**2.2. Pressure-Driven Electronic Topological Transition**

The magnetization curves were measured at 5 K to further investigate the magnetic behavior of the martensite phase. A sudden drop ≈0.6 GPa was observed during loading (**Figure 4**a,b). To understand this anomalous nonlinear behavior, we performed the first-principles calculations. Firstly, we obtained highly comparable values of saturation magnetization with the experimental measurements. Interestingly, a synchronous sudden drop was also observed at the same critical pressure during loading, which reproduces the experimental behavior (Figure 4b). This decrease in saturation magnetization implies a change in the band structure near the Fermi level under the pressure.

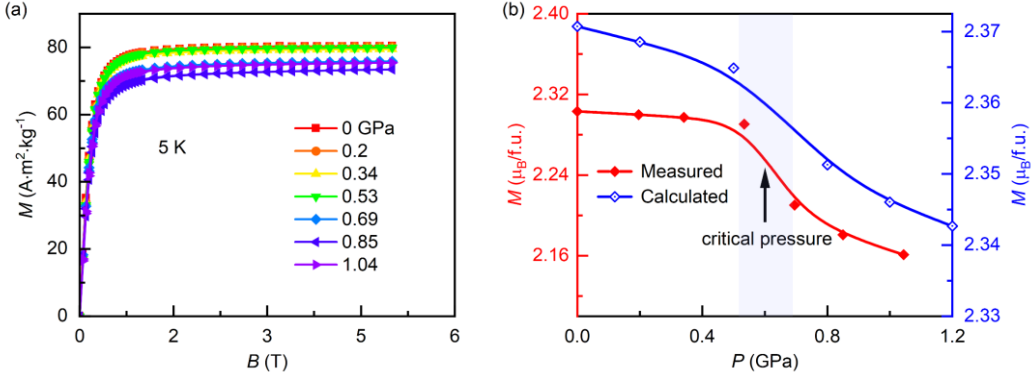

**Figure 4.** a) Saturation magnetization of martensite phase at 5 K. b) Pressure-dependent saturation magnetization obtained from (a) (red solid diamond) and first-principles calculations (blue hollow diamond).

As the saturation magnetization is determined by the difference between the integrals of spin-up and spin-down channels of DOS, we further revealed the underlying mechanism of the anomalous behavior by analyzing the DOS evolution, as shown in **Figure 5**a. It is possible to calculate the desired electronic structure by building a supercell with a chemical formula of $Mn_{0.5}Fe_{0.5}NiGe_{0.5}Si_{0.5}$, which is very close to the composition of the studied sample. In the spin-up channel, there is a DOS peak around the Fermi level at the ground state. A sudden change of the peak location can be clearly seen ≈0.6 GPa, with a sudden drop of DOS values around the Fermi level (Inset of Figure 5a). An orbital level experiences a shift to higher energy at ≈0.6 GPa, as depicted in the Figure 5b. When the pressure is lower than 0.6 GPa, an energy band of d orbital is located at the Fermi level, producing a DOS peak at the Fermi level. As the pressure approaches 0.6 GPa, the energy band leaves the Fermi level and exhibits a right shift on the DOS at the Fermi level. This d orbital corresponds to an electronic Fermi pocket around the Fermi level, and the right shift of this orbital indicates the disappearance of the electronic Fermi pocket at the Fermi level. According to the experimental results and theoretical calculations, the evolution of DOS/band structure around the Fermi level in Figure 5 reveals an ETT[17] ≈0.6 GPa. In Figure 5a, a peak of the spin-down DOS synchronously has a significant left shift. The sudden change in the energy band at the Fermi level ≈0.6 GPa leads to a change in the number of spin-up and spin-down electronic states, resulting in a decrease in the saturation magnetization (Figure 4b). This is the first observation of pressure-induced ETT in martensitic systems by theoretical calculations and experimental measurements. During the ETT, the electronic structure, including the DOS and Fermi surface, undergoes changes, resulting in a decrease in magnetic moment.

With the shift of the orbital level under pressure, the lattice parameters of the system also show sudden pressure-dependent changes. We plot the pressure dependence of the axes in Figure 5c. While one can see a linear decrease of the *a*-axis, the *b*-axis shows a sudden expansion and *c*-axis a sudden shortening, resulting in a shrink of the unit cell, as shown in Figure 5d. There is an anisotropy in the pressure-driven changes of the lattice parameters, leading to a decrease in orthorhombicity of the crystal unit cell during the ETT.

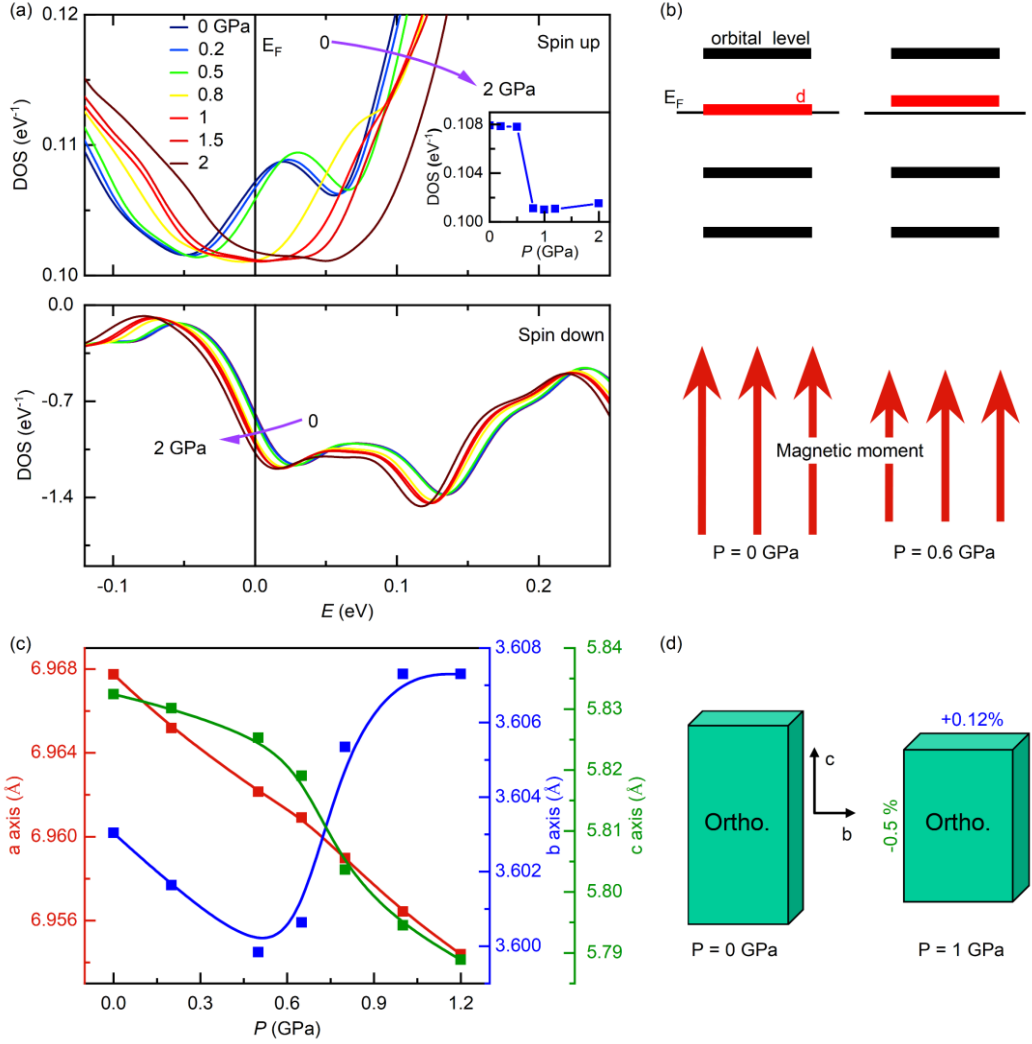

**Figure 5.** a) Pressure dependent DOS/band evolutions of $Mn_{0.5}Fe_{0.5}NiGe_{0.5}Si_{0.5}$, revealing a sudden right-shift of an energy orbit ≈0.6 GPa. The inset shows the DOS of spin-up channel evolving with hydrostatic pressure at the Fermi level. b) shows the diagram of band structure and magnetic moment evolution with pressure. c,d) Pressure-dependent lattice parameters, showing a remarked jump in $b$ and $c$-axes, with $a$-axis and volume weak change.

## 3. Conclusion

In conclusion, we studied the effect of hydrostatic pressure on the MPT and magnetic properties of magnetic martensitic $Mn_{0.54}Fe_{0.46}NiGe_{0.4}Si_{0.6}$. Due to the large driving efficiency ($dP/dT_t$), the material shows a considerable BCE. The MCE with a large magnetic entropy change (-30 J·kg$^{-1}$·K$^{-1}$ in a field change of 5 T) was modulated from 339 to 273 K by hydrostatic pressure, demonstrating potential for solid-state cooling in a wide temperature range. Additionally, a sudden drop in magnetic moment at 0.6 GPa was observed experimentally. The DOS and lattice parameters of low-temperature martensitic phase were calculated by the first-principles calculations, and the sudden change of DOS at the Fermi level and lattice parameters can be observed theoretically. The calculated trend of saturation magnetization with hydrostatic pressure is in good agreement with our experiments. This work studied the effect of hydrostatic pressure on BCE, MCE and ETT, which not only provides a material platform near room temperature for refrigeration application but also gives the evidence of the correlation between saturation magnetization and DOS around the Fermi level under hydrostatic pressure.

## 4. Experimental Section

Sample preparation

Arc melting in argon atmosphere was used to melt the polycrystalline ingots. The ingots were sealed in the quartz tubes and annealed at 1123 K for 5 days, then cooled slowly. The MPT was characterized by magnetic properties.

DFT Calculations

Lattice parameters, magnetic moment, DOS, ELF and SPD under pressure were calculated for $Mn_{0.5}Fe_{0.5}NiGe_{0.5}Si_{0.5}$ based on density functional theory (DFT) using the pseudopotential method with plane-wave-basis set.[29-31] Pseudopotentials with a planewave-basis cutoff of 550 eV and a k-mesh of 10×5×6 were used within the generalized gradient approximation (GGA), following the PBE parametrization scheme. The geometry optimizations for the atomic site occupancy in the cell were performed using the Broyden–Fletcher–Goldfarb–Shanno minimization scheme.[32] The break condition of the total energy and the ionic relaxation loop for relaxation were set to $5\times10^{-7}$ eV and $5\times10^{-6}$ eV, respectively. Then, vaspkit[33] and vesta[34] were used for post-processing of ELF and SPD.

High-Pressure magnetization measurements

The MPT was characterized by magnetic properties. The magnetic measurements were performed by Quantum Design Superconducting Quantum Interference Device-Vibrating Sample Magnetometer (SQUID-VSM) with a pressure cell. The pressure transmitting medium Daphne 7373 was used. The pressure was calibrated by the critical temperature of superconducting transition of pure Pb.[35]

**Acknowledgements**

J.Y. and X.L. equally contributed to this work. This work was supported by the National Key R&D Program of China (2019YFA0704900, 2022YFA1403400), the Fundamental Science Center of the National Natural Science Foundation of China (No. 52088101), the Strategic Priority Research Program (B) of the Chinese Academy of Sciences (CAS) (No. XDB33000000), the Synergetic Extreme Condition User Facility (SECUF), and the Scientific Instrument Developing Project of CAS (No. ZDKYYQ20210003). XXZ acknowledges the financial support from King Abdullah University of Sciences and Technology (KAUST) under Award Nos ORA-CRG10-2021-4665 and ORA-CRG11-2022-5031.